\documentclass[pre,preprint,onecolumn,amssymb,showkeys,superscriptaddress,notitlepage]{revtex4-1}
\pdfoutput=1
\usepackage{graphicx}
\usepackage{color}
\usepackage{epsfig}
\usepackage{latexsym}
\usepackage{bm}
\usepackage{ulem}
\usepackage{color}
\usepackage{enumerate}
\usepackage{enumitem}

\renewcommand{\figurename}{{Figure}}
\usepackage{dcolumn}
\usepackage{amsmath,amssymb}    
\usepackage{bm}
\usepackage{hyperref}
\usepackage{latexsym}
\usepackage{color}
\usepackage{subfigure}
\usepackage{fancyhdr}
\usepackage{chngcntr}

\def\beq{\begin{equation}}
\def\eeq{\end{equation}}

\def\beq{\begin{equation}}                           
\def\eeq{\end{equation}}
\def\bea{\begin{eqnarray}}                           
\def\eea{\end{eqnarray}}

\usepackage{lineno}

\begin{document}

\title{How enzymatic activity is involved in chromatin organization}


\author
{\parbox{\linewidth}{\centering 
Rakesh Das$^{1\ast}$, Takahiro Sakaue$^{2}$, G. V. Shivashankar$^{3,4}$, Jacques Prost$^{1,5}$ and \\ Tetsuya Hiraiwa$^{1,\ast}$\\
\normalsize{$^{1}$Mechanobiology Institute, National University of Singapore, Singapore}\\
\normalsize{$^{2}$Department of Physics and Mathematics, Aoyama Gakuin University, Kanagawa, Japan}\\
\normalsize{$^{3}$ ETH Zurich, Zurich, Switzerland.}\\
\normalsize{$^{4}$ Paul Scherrer Institute, Villigen, Switzerland.}\\
\normalsize{$^{5}$Laboratoire Physico Chimie Curie, Institut Curie, Paris Science et Lettres Research University, CNRS UMR168, Paris, France}\\
\normalsize{$^\ast$To whom one may correspond :mbird@nus.edu.sg; mbithi@nus.edu.sg.}
}
}

\date{\today}


\begin{abstract}
Spatial organization of chromatin plays a critical role in genome regulation. 
Various types of affinity mediators and enzymes have been attributed to regulate spatial organization of chromatin from a thermodynamics perspective. 
However, at the mechanistic level, enzymes act in their unique ways. 
Here, we construct a polymer physics model following the mechanistic scheme of Topoisomerase-II, an enzyme resolving topological constraints of chromatin, and investigate its role on interphase chromatin organization. 
Our computer simulations demonstrate Topoisomerase-II’s 
ability to phase separate chromatin into eu- and heterochromatic regions with a characteristic wall-like organization of the euchromatic regions. 
Exploiting a mean-field framework, we argue that 
the ability of the euchromatic regions crossing each other due to enzymatic activity of Topoisomerase-II induces this phase separation. 
Motivated from a recent experimental observation on different structural states of the eu- and the heterochromatic units, we further extend our model to a bidisperse setting and show that the characteristic features of the enzymatic activity driven phase separation survives there.
The existence of these characteristic features, even under the non-localized action of the enzyme, highlights the critical role of enzymatic activity in chromatin organization, and points out the importance of further experiments along this line.
\end{abstract}


\maketitle


\section{Introduction}
During interphase, chromatin in a nucleus is densely packed and unable to move freely around the nucleus, resulting in a highly constrained positioning of genes.
Nowadays, it is acknowledged that such physical spacing of chromatin (genes) is critical in regulating biochemical and transcriptional abilities of genes \cite{ShivashankarNatRev2017, WangCell2018, ElginColdSpringHarbPerspecBiol2013},
and proper functionality of the genomic content depends on the nonrandom organization of chromatin \cite{SoloveiCurrOpinCellBiol2016, DekkerTrendBiochemSci2020}.
Three dimensional contact mapping techniques have revealed that chromatin is compartmentalized into euchromatic (EC) and heterochromatic (HC) regions 
\cite{LiebarmanAidenScience2009, NicodemiNatMethods2021}. 
In the EC regions, the nucleosomes are widely separated allowing greater access of the embedded genes to various regulatory factors, and therefore, EC regions are transcriptionally active. 
In contrast, HC regions comprise densely packed nucleosomes, and they are transcriptionally repressed. 
Recent literature \cite{LarsonNature2017, LarsonBiochem2018, StromNature2017, GibsonCell2019, ErdelBiophysJ2018, DekkerTrendBiochemSci2020} have argued phase separation as one of the driving mechanisms for such compartmentalization of chromatin.
Affinity among HC regions, mediated by a diverse range of molecular agents \cite{ErdelBiophysJ2018, DekkerTrendBiochemSci2020}, are believed to induce such phase separation in chromatin. 
Besides this affinity-induced phase separation, many active agents (which are ATP dependent and therefore capable of driving the system out-of-equilibrium) play crucial roles in chromatin organization, e.g., extruder-motor assisted loop formation \cite{MirnyPNAS2018, MirnyCurrOpinCellBiol2019} or RNA polymerase II mediated transcriptional pocket formation \cite{JulicherNatComm2021}.

Nuclear media is full of various types of affinity mediators and active agents.  
To investigate how those agents affect chromatin organization, it can be useful to employ concepts of physics.
As a matter of fact, polymer physics models have been successfully employed to explain various aspects of experimental observations \cite{LiebarmanAidenScience2009, MirnyFEBSLett2015, NicodemiNatMethods2021}.
Modeling chromatin as block copolymers and tuning the affinity among those blocks could reproduce the plaid-like pattern observed in contact maps \cite{JostNAR2014,  MartinNature2019, SpakowitzPNAS2018}. 
Here the blocks represent genomic regions with different epigenetic marks, e.g., H3K9ac and H3K27me3 histone marks characterizing EC and HC regions, respectively.
Polymer physics approach has been also useful to implicate the role of active biophysical processes on chromatin organization \cite{KremerPRL2017, GautamNAR2014, GautamBiophysJ2020}. 
By modeling active sites of active agents as local regions at higher temperatures as compared to the rest of the media, these studies highlighted the effect of out-of-equilibrium processes on chromatin organization. 
However, 
at the mechanistic level it is likely that the activity of each enzyme will affect dynamics beyond just
effective-temperature inhomogeneity. 
We need dedicated studies to elucidate how the enzymatic activity can affect the microphase separation (MPS) structures beyond just a thermodynamics phenomenology by employing the mechanistic model focusing on a specific type of enzyme.

In this paper, we focus on topoisomerase enzyme of type II (Topo-II), an active agent that plays a pivotal role in resolving topological constraints of chromatin which emerge due to dense packing \cite{NitissNatRevCancer2009, VologodskiiPhysLifeRev2016, RocaNuclAcidRes2009, NitissNatRevMolCellBiol2016, KouzineTranscription2013, HsiehAnnuRevBiochem2013, PoljakTrendCellBiol1995}, and investigate the effects of this enzyme on chromatin organization.
Topo-II transports one DNA duplex across another, which is cleaved transiently and resealed after transport.
The role of this enzyme in processes like transcription, replication, and segregation of sister chromatids has been investigated extensively \cite{NitissNatRevCancer2009, NitissNatRevMolCellBiol2016, RosenfeldScience2006}; here, we investigate the possibility for this enzyme to modify chromatin organization during interphase.
To accomplish this aim, we 
developed 
an active polymer model mimicking the mechanistic scheme of Topo-II's activity.
We find that Topo-II has inherent ability to induce MPS in chromatin. 
Using simplified model studies, we argue that the underlying mechanism of Topo-II driven phase separation is of a new type; the effective 
phantomness of polymer segments (i.e., the ability of the segments crossing each other)
due to Topo-II activity induces phase separation. 
We find that Topo-II induces a characteristic `wall'-like structure of EC regions - a feature that has not been observed in other models studying phase separation of chromatin. 
Further, we investigate how such MPS structure is affected by bidispersity of the chromatin.
The idea of considering the case of a bidisperse chromatin is inspired from ref.~\cite{XuCellRep2018}, which showed that epigenetic marks associated with EC and HC regions remains as clusters of {\it different sizes}.

\section{Results} \label{results}
\begin{figure}[t]
\includegraphics[width=\linewidth]{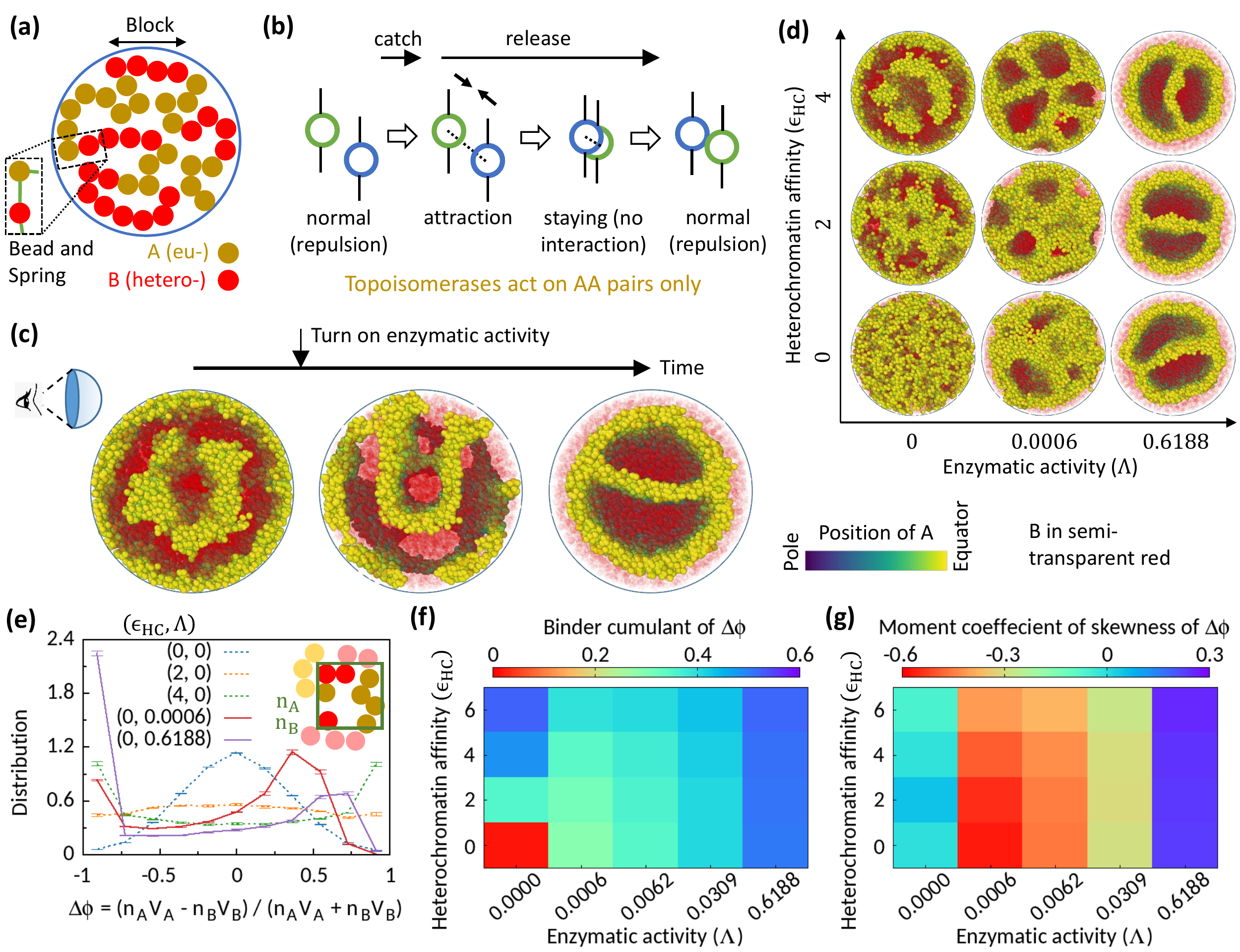}
\caption{ \footnotesize
{\bf Microphase separation of eu- and heterochromatic regions due to enzymatic activity.}
(a) A random multiblock copolymer comprising A and B beads connected by springs confined within a spherical cavity. All the data are shown for block size $b=4$.
(b) Topo-II enzyme catches two A's in spatial neighbourhood. Through a sequence of processes, it passes one A across another with some probability and eventually releases both A's.
(c) Sample snapshots (hemisphere cuts) showing that microphase separation configuration changes significantly after turning on enzymatic activity. The color bar indicates position of A's, and B's are shown in semi-transparent red. Parameters used\textemdash $\epsilon_{HC}=4$ and $\Lambda=0.6188$.
(d) Sample snapshots showing microphase separation in response to heterochromatin affinity and enzymatic activity.   
(e) Inset\textemdash The cavity is divided into small grids, and $n_A$ and $n_B$ stand for the numbers of the respective beads within individual grid.
Main\textemdash $V_{A,B}$ represent volume of the respective beads.
Distribution of $\Delta\phi$ goes from unimodal to bimodal as the system phase separates. 
Time-averaged data shown, and error bars indicate standard deviations over realizations.
(f) Binder cumulant $1-\langle(\Delta\phi)^4\rangle_P/3\langle(\Delta\phi)^2\rangle_P^2$ value greater than zero indicates deviation of the $\Delta\phi$-distribution from the Gaussian profile.
(g) Moment coefficient of skewness $E\left[ \left( \Delta\phi - \langle\Delta\phi\rangle \right)^3 \right] / \left\{ E\left[ \left( \Delta\phi - \langle\Delta\phi\rangle \right)^2 \right] \right\}^{3/2}$ captures the  asymmetry in the $\Delta\phi$-distribution about its mean in the presence of Topo-II.
}
\label{Fig1}
\end{figure}

\noindent{\bf Polymer model of Topoisomerase's activity.}
We developed a copolymer model to study three-dimensional organization of a $50.807$ Mbp long chromatin confined within a spherical cavity of diameter $1.4112$ $\mu$m. The copolymer comprises two types of equal-sized beads, A and B, connected by springs (Fig.~\ref{Fig1}a). These beads represent EC and HC contents, respectively, each mirroring $2.4203$ kbp of chromatin (see Methods). Each type of bead appears in blocks of size $b$, and the blocks are distributed randomly along the polymer. 
Affinity among HC regions is modeled by a short range attraction between B beads, parametrized by $\epsilon_{HC}$ \cite{ErdelBiophysJ2018, DekkerTrendBiochemSci2020, MartinNature2019, MirnyPNAS2018}.

Topo-II relaxes topological constraints of a chromatin in a {\it catch-and-release} mechanism\textemdash it catches two DNA segments in spatial proximity, and through a sequence of processes including ATP hydrolysis, it eventually transports one DNA segment across the other and releases both segments \cite{RocaNuclAcidRes2009, NitissNatRevCancer2009}. 
We engineered our polymer model in a particular way to mimic this catch-and-release mechanism of Topo-II's activity.  
First, a Topo-II catches two beads in spatial proximity with a Poisson rate $\lambda_{ra}$ (Fig.~\ref{Fig1}b). 
The beads bound to the enzyme no more exert steric repulsion to each other; instead they attract each other.  
This attraction state mimics the locked N-gate state of Topo-II that brings two DNA segments closer to each other \cite{RocaNuclAcidRes2009}.
Next, the attraction between those two beads is turned off with a rate $\lambda_{an}$, and the beads stay there for a while without any steric interaction among themselves. This step allows the beads to pass across each other stochastically. Eventually, the enzyme unbind from the beads with a rate $\lambda_{nr}$, and the beads return to their normal state with steric repulsion between themselves. 
We define enzymatic activity as $\Lambda = \lambda_{ra}\left( 1/\lambda_{an} + 1/\lambda_{nr} \right)$, which can be tuned in experiments by controlling ATP concentration \cite{WangJBiolChem1993}.

Experiments using budding yeast suggests that Topo-II mainly works on the nucleosome free regions of the genome \cite{GrunsteinPNAS2011, KouzineTranscription2013}. 
As it is more likely for Topo-II to find nucleosome free bare DNA segments in the EC regions, we assume that Topo-II works on AA pairs only. 
Hereafter, we refer to this polymer model as the {\it monodisperse differential active model} (MdDAM). We simulate this model using Brownian dynamics at physiological temperature $310$ K (see Methods). 
The composition of the copolymer system is quantified by volume fraction $\phi_A$ of A beads, defined as the ratio of the total volume of A beads to that of all the beads. \\


\noindent{\bf Topoisomerase affects chromatin organization.}
To investigate the role of Topo-II on chromatin organization, we compare the morphology of chromatin organization in the absence and the presence of enzymatic activity. 
We start our simulation in the absence of Topo-II but for finite HC affinity ($\epsilon_{HC}>2$) and observe MPS of chromatin into EC-rich and HC-rich domains (Fig.~\ref{Fig1}c-left). 
Interestingly, once the enzymatic activity is turned on during the simulation, the MPS structure evolves into a significantly different morphology (Fig.~\ref{Fig1}c-centre and right). 
This suggests the importance of Topo-II's activity in chromatin organization. 

Next, we focus on the differences between the MPS induced by HC affinity and that due to Topo-II's activity. In the absence of Topo-II, the chromatin microphase separates in response to HC affinity, as shown in Fig.~\ref{Fig1}d for symmetric composition ($\phi_A=0.5$). To quantify this MPS, we define an order parameter $\Delta\phi = \left(n_AV_A-n_BV_B\right)/\left(n_AV_A+n_BV_B\right)$, where $V_{A,B}$ represents the volume of the respective beads, and $n_{A,B}$ represents the number of the corresponding beads in a coarse-grained locality. The system remains in a disordered state in the control case ($\epsilon_{HC}=0$), and the distribution $P\left(\Delta\phi\right)$ (see Methods) of the order parameter follows a Gaussian curve (Fig.~\ref{Fig1}e). However, $P\left(\Delta\phi\right)$ flattens and eventually becomes bimodal with HC affinity, suggesting the appearance of EC-rich and HC-rich domains. This phenomenology is well captured in the phase diagram shown in Fig.~\ref{Fig1}f, where the Binder cumulant increases with $\epsilon_{HC}$ suggesting deviation of $P\left(\Delta\phi\right)$ from the Gaussian distribution.

Interestingly, Topo-II not only alters chromatin organization, but 
alone can drive MPS of chromatin (Fig.~\ref{Fig1}d). 
In the presence of Topo-II, $P\left(\Delta\phi\right)$ becomes bimodal (Fig.~\ref{Fig1}e), which signifies Topo-II's ability to induce MPS. 
However, the bimodal profile of $P\left(\Delta\phi\right)$ is strongly asymmetric about its mean in the presence of enzymatic activity,
as shown in the distribution (Fig.~\ref{Fig1}e) and quantified by skewness (Fig.~\ref{Fig1}g).
This is the stark difference from the affinity-induced case, where 
 we have a symmetric profile of order parameter distribution around $\Delta\phi=0$ (Fig.~\ref{Fig1}e, g), 
as expected for the symmetric composition.
This suggests that the phase separation induced by the Topo-II activity attributes to a fundamentally new mechanism that is qualitatively distinct from the affinity-induced case. \\

\begin{figure}[t]
\includegraphics[width=\linewidth]{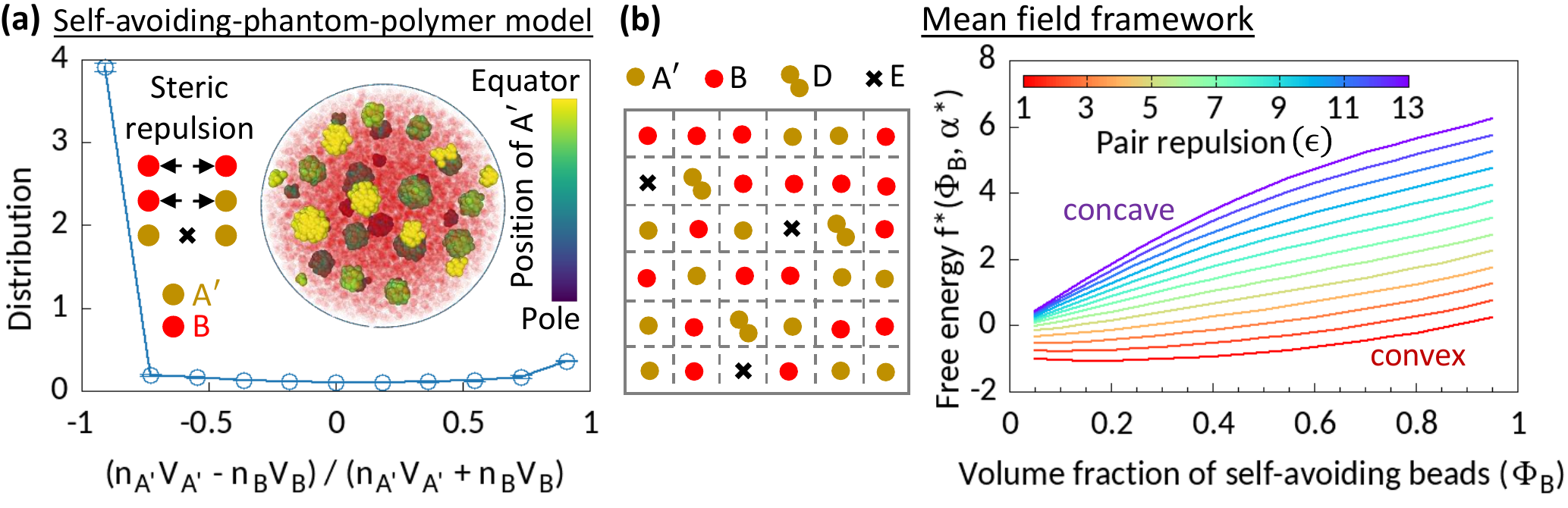}
\caption{ \footnotesize
{\bf Phase separation in system comprising self-avoiding and phantom regions.}
(a) An equilibrium copolymer system comprising phantom (A$^\prime$) and self-avoiding (B) beads is simulated in the absence of steric interaction between A$^\prime$’s. The system shows microphase separation. 
A sample snapshot (hemisphere cut) is shown where B's are in semi-transparent red. 
Time-averaged data shown for the distribution, and error bars indicate standard deviations over realizations.
(b) Left\textemdash Schematic of a lattice space filled with phantom (A$^\prime$) and self-avoiding (B) beads. A$^\prime$ beads can form doublets (D) resulting empty (E) sites. Mean field calculation gives an effective attraction among B’s. Right\textemdash Free energy $f^*$ curves drawn for critical doublet fraction $\alpha^{*}\left(\Phi_B,\epsilon\right)$. $f^*$ shows convex to concave transition with pair repulsion parameter $\epsilon$, suggesting a phase separation in the system. 
}
\label{Fig2}
\end{figure}

\noindent{\bf Plausible mechanism of enzymatic activity driven phase separation.}
To understand the underlying physical mechanism of enzymatic activity driven phase separation, we reduce MdDAM to its equilibrium version. 
We replace the self-avoiding A beads of MdDAM by phantom A$^\prime$ beads such that there is no steric interaction between A$^\prime$A$^\prime$ pairs at any time.   
Interestingly, this simplified model, called a self-avoiding-phantom-polymer model, shows microphase separation even in the absence of HC affinity (Fig.~\ref{Fig2}a). 
A$^\prime$'s prefer other A$^\prime$'s as their neighbors because that saves the steric energy cost of the system. 
Moreover, the number of available 
microstates
and therefore the entropy of the system increases if A$^\prime$'s stay close to each other. 
We argue that both these energetic and entropic advantages drive microphase separation in this equilibrium polymer system. 
From a physics standpoint, it would be interesting 
to construct a mean-field framework for the self-avoiding-phantom-polymer model in the spirit of ref.~\cite{FredricksonMacromolecules1992}; however, that is a non-trivial task and beyond the scope of the current article. 
Instead, we construct a relevant but reduced mean-field framework by relaxing the polymeric degrees of freedom of the system.
In this simplification, the
MPS morphology observed for the self-avoiding-phantom-polymer model will vanish, 
but the fundamental mechanism driving the phase separation should still be at work. We discuss this simplified mean-field framework in the next paragraph.

We consider a lattice system filled with A$^\prime$ and B beads (Fig.~\ref{Fig2}b). 
Two A$^\prime$ beads can form a doublet (D), resulting in an empty lattice site (E). 
Given a doublet fraction $\alpha \in [0,1/2]$, our mean field level calculation gives a term $\epsilon c(\alpha) \Phi_B^2$ in the free energy density, where $\Phi_B$ is the volume fraction of B's in this mean-field model, and the coefficient $c(\alpha)<0$ (see Methods). 
This term suggests that the phantomness of A$^\prime$ eventually induces an effective attraction between B's, driving a phase separation in the system.
The observed convex to concave transition of the profile of the reduced free energy density $f^*$, expressed for the critical doublet fraction $\alpha^*$, with pair repulsion parameter $\epsilon$ justifies our claim (Fig.~\ref{Fig2}b).

A's in MdDAM transiently behave like phantom beads. Seeing how phantomness of one type of beads can induce a phase separation in the simplified model systems described above, we argue that a similar physical mechanism is responsible for the enzymatic activity driven phase separation.
However, we must note that the wall-like organization of A beads in the Topo-II driven microphase separated configurations, as elaborated in the next section, is a unique feature not observed in the phase separation between self-avoiding and phantom segments (Fig.~\ref{Fig2}a).

The possibility of the phase separation by the transient attraction state during enzymatic activity was ruled out by simulating a representative copolymer system with comparable attraction strength among A beads (Supplementary Fig.~1). 

In the snapshots shown for $\Lambda=0.6188$ in Fig.~\ref{Fig1}d, the phase separation structure is not affected by the HC affinity.
Probably the energy cost derived from the entropic advantage due to Topo-II's activity might overpower HC affinity, resulting in this robust phase separation structure for strong enzymatic activity. 
\\

\begin{figure}[t]
\includegraphics[width=\linewidth]{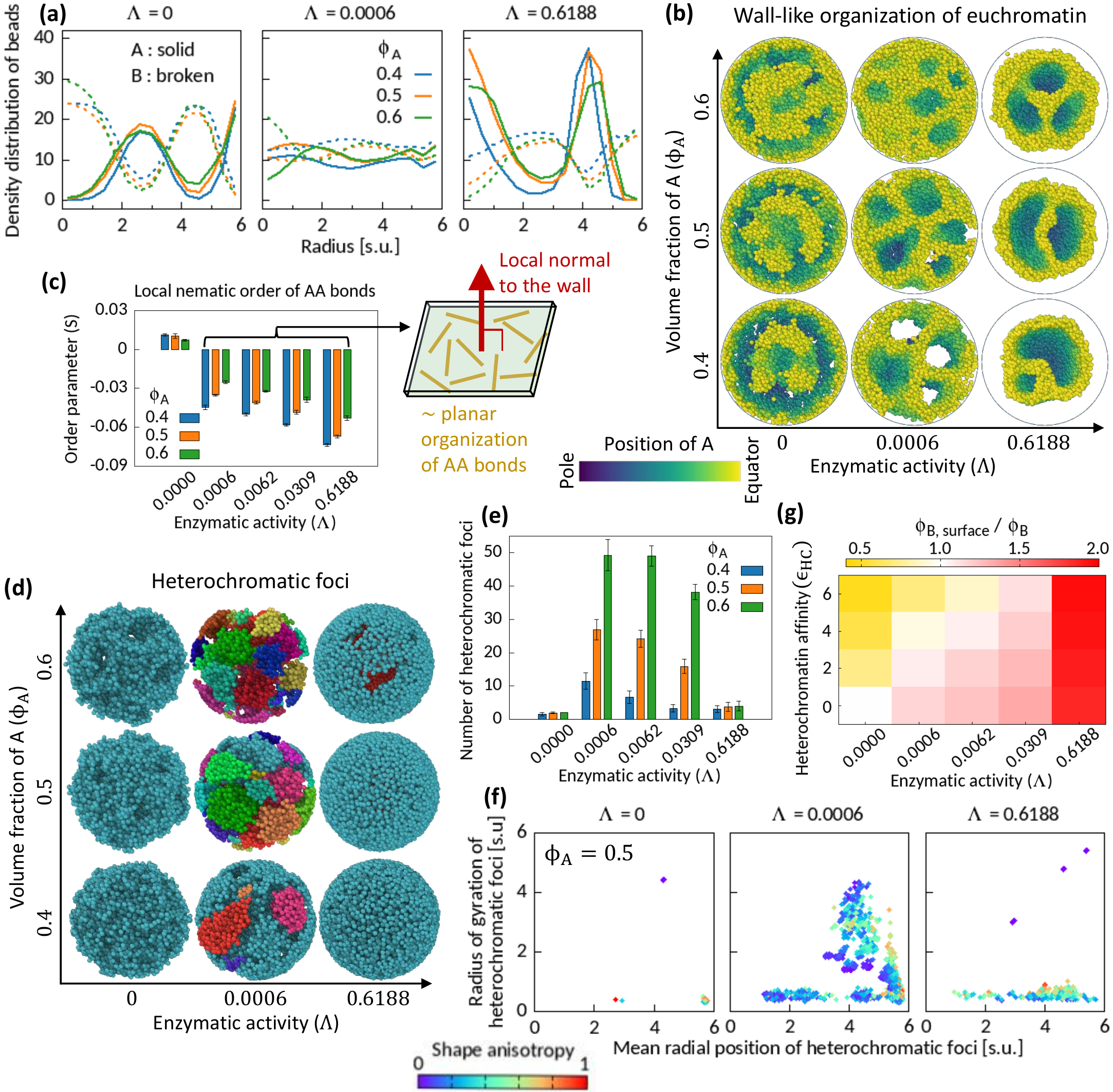}
\caption{ \footnotesize 
{\bf Characteristics of Topo-II induced microphase separation configurations.}
(a) Density distribution of A and B beads in radial direction, plotted for fixed $\epsilon_{HC}=4$. 
(b) Sample snapshots (hemisphere cuts) showing wall-like organization of A's for $\Lambda > 0$.
(c) Local nematic order parameter of AA bonds. Schematic shows approximate organization of AA bonds in the wall.
(d\textendash f) Heterochromatic foci features.
Sample snapshots (d) and number (e) of heterochromatic foci are shown.
Different color in (d) indicates distinct focus.
Time-averaged data are shown in (e) and the error bars indicate standard deviations over realizations.
(f) Position and size of individual focus are respectively represented by the mean radial coordinates of the member-B's and the radius of gyration of the focus. Shape anisotropy ranges from zero to unity for spherical and line-shaped foci, respectively. 
(g) Volume fraction of B's at the surface over the global volume fraction of B is shown in $\Lambda$-$\epsilon_{HC}$ space for $\phi_A=0.5$. Time-averaged data are shown. 
}
\label{Fig3}
\end{figure}

\noindent{\bf Wall-like organization of EC due to Topo-II.}
We examine the obtained microphase separated snapshots to understand the effect of Topo-II on chromatin organization. 
The number density of the beads suggests an alternating and complimentary organization of A and B beads along the radius of the cavity (Fig.~\ref{Fig3}a). 
Interestingly, in the presence of enzymatic activity, we note a wall-like appearance of EC domains (Fig.~\ref{Fig3}b). 
By wall, we mean that the spread of EC regions along a (curvilinear) plane is broader than that along its normal direction. 
This is a robust feature we note in all of our simulated configurations in the presence of enzymatic activity. 
We can discern this feature clearly from the A-only snapshots shown for $\Lambda=0.6188$ in Fig.~\ref{Fig3}b. 
Predominantly, the A's form a wall-like spherical shell, and as per the given composition of the system, the rest of the A's too arrange themselves in wall-like manner.
This kind of organization is in sharp contrast with surface-minimizing globule shaped organization of beads in response to affinity driven phase separations. 

Next, we examine the orientation of the chromatin segments in the  wall-like organization of the EC regions by measuring a local nematic order $S$ of the AA bonds. 
Within coarse-grained localities, the AA bonds show negative nematic order parameter in the presence of the enzymatic activity, while no significant order is observed for $\Lambda=0$ (Supplementary Fig.~2, also see Methods). 
The negative nematic order of AA bonds implies that those bonds are approximately parallel to the plane along the wall \cite{PGdGennesBook}. 
We calculate the mean 
$S$ of the AA bonds, averaged across the cavity except near the surface (see Methods), and plot it in Fig.~\ref{Fig3}c.
The consistent trend of negative $S$ for $\Lambda \neq 0$ portrays the association of the characteristic wall-like organization and the local order of the AA bonds therein.
\\


\noindent{\bf Effect of Topo-II on heterochromatic foci.}
For the cell to function properly, the number, size, and spatial position of heterochromatic foci has to be critically regulated \cite{FodorAnnuRevCellDevBiol2010, WangCell2018, ElginColdSpringHarbPerspecBiol2013}.
So, we segmented heterochromatic foci (see Methods) from the simulated snapshots and investigate their features to understand the role of Topo-II's activity on them. 
We show sample snapshots of segmented foci on Fig.~\ref{Fig3}d.  
Most of the B's remain connected under the action of HC affinity and in the absence of enzymatic activity (Fig.~\ref{Fig3}d). Consequently, we count a small number of foci (Fig.~\ref{Fig3}e)\textemdash a relatively big sponge-like focus spread across the cavity, and a few other small foci scattered elsewhere (Fig.~\ref{Fig3}f).
For strong enzymatic activity, we count a small number of foci with mainly two dominating modes\textemdash one, localized near the surface of the cavity having various sizes depending on $\phi_A$, and second, a focus localized inside the cavity having a notable morphology (Fig.~\ref{Fig3}d\textendash f and Supplementary Fig.~4). 
For moderate enzymatic activity, we see many foci of various shapes and sizes (Fig.~\ref{Fig3}d\textendash f and Supplementary Fig.~3). The scatter plot in Fig.~\ref{Fig3}f is color coded by the shape anisotropy (see Methods) of the individual foci, and it suggests the appearance of foci with various shapes. \\


\noindent{\bf Topo-II in determining surface profile of chromosome territory.}
In conventional nucleus, HC regions accumulate near the nuclear membrane, whereas transcriptionally active genes mostly localize at the intermingling regions of two chromosome territories \cite{MartinNature2019, ShivashankarNatRev2017, ShivashankarCurrOpinCellBiol2019}.
Thus, there exists an orchestration of mechanisms that determines whether EC and/or HC regions will localize at the surface of a chromosome territory. 
While a combination of strong HC affinity \cite{LarsonNature2017, StromNature2017}, and interaction between lamina and HC contents \cite{CollasGenomeBiolRev2020, SteenselNature2008} can explain this phenomenon, one cannot exclude the possibility of other mechanisms playing a significant role in this regard. 
We note that the enzymatic activity of Topo-II competes with HC affinity in determining surface localization profile. 
HC affinity pulls B's inward the cavity to minimize the interfacial energy cost. 
On the other hand, enzymatic activity drives A's inside the cavity. 
To illustrate this competition, we calculate the ratio of the mean volume fraction of B's at the surface of the cavity ($\phi_{B, surface}$, see Methods) to the global volume fraction of B's, $\phi_B=1-\phi_A$, and construct a heatmap (Fig.~\ref{Fig3}g).
The heatmap  manifests this competition and hints at the existence of an isoline on the $\Lambda$-$\epsilon_{HC}$ plane where $\phi_{B, surface} = \phi_B$. 
In Fig.~\ref{Fig3}g, we show the heatmap for the symmetric composition; however, we find similar heatmap for other $\phi_A$'s.
\\

\begin{figure}[t]
\includegraphics[width=\linewidth]{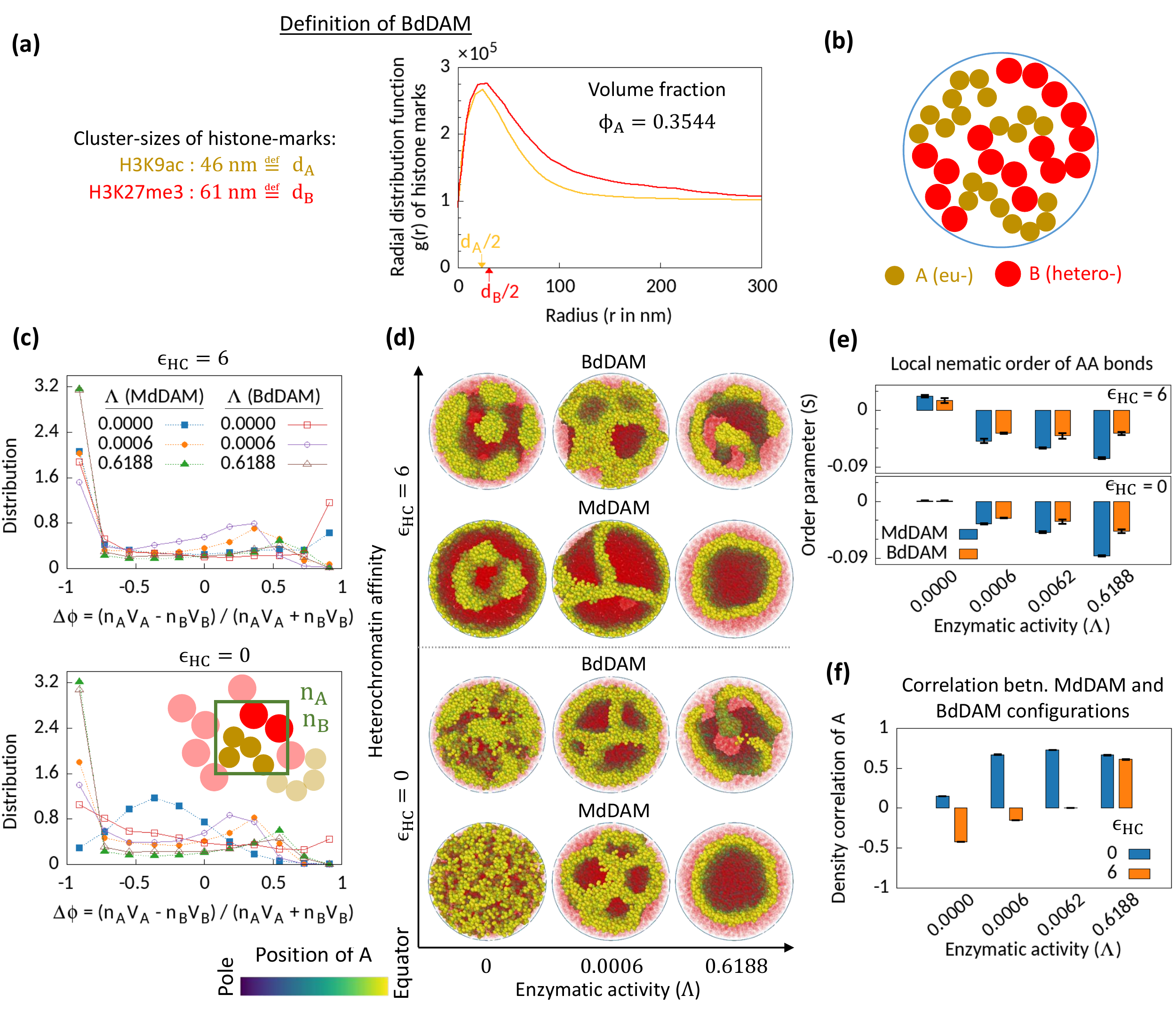}
\caption{ \footnotesize
{\bf Microphase separation in bidisperse model, motivated from super-resolution microscopy data.}
(a) Extracted data for mean cluster sizes and radial distribution functions of histone marks characteristic to eu- and heterochromatic regions. Following the experimental data, we set 
volume fraction $\phi_A=0.3544$.
(b) Schematic of the bidisperse random multiblock copolymer model.
(c\textemdash f) Comparison of the bidisperse model (BdDAM) with the monodisperse model (MdDAM).
Time-averaged $\Delta\phi$-distributions are shown in (c), and the error bars over realizations are not shown as those are smaller than the symbol sizes.
(d) Sample snapshots (hemisphere cuts) are shown, where the B's are represented in semi-transparent red. The bidisperse system shows phase separation even for $\epsilon_{HC}=0$ and $\Lambda=0$. 
(e) Local nematic order parameters, averaged over realizations, are shown and the error bars indicate the corresponding standard deviations.
(f) Cross correlation of local density of A's between MdDAM and BdDAM configurations are shown (see Methods for definition).
The data shown are averaged over realizations, and the error bars indicate the corresponding standard deviations.
}
\label{Fig4}
\end{figure}

\noindent{\bf Bidisperse chromatin model.} 
A recent super-resolution microscopy study showed that at the epigenetic level, histone marks characterizing EC and HC regions remain at different structural states \cite{XuCellRep2018}.
Active histone marks like H3K9ac 
form distinct and small clusters as compared to condensed large aggregates of the repressive marks such as H3K27me3. 
An implication of this experimental observation from the polymer physics model perspective is that the beads representing EC and HC regions have different sizes. Thus we came up with the idea of bidisperse chromatin. 
Bidispersity is known to affect the phase separation pattern of colloidal systems \cite{HueckelNatRevMat2021, AsakuraJChemPhys1954, VrijPAC1976}. 
Here, we study the effect of enzymatic activity on bidisperse chromatin.
We modify our copolymer model in such a way that the A and B bead sizes are different from each other and respectively equal to the mean sizes of H3K9ac- and H3K27me3-clusters (Fig.~\ref{Fig4}a\textendash b).
Hereafter, we refer this modified setting as {\it bidisperse differential active model} (BdDAM). 
To simulate this system for a biologically relevant composition, we extracted the radial distribution function data from ref.~\cite{XuCellRep2018}, and obtained the corresponding 
volume fraction as $\phi_A=0.3544$ (Fig.~\ref{Fig4}a; also see Methods). 

We first simulate BdDAM without any HC affinity and enzymatic activity and obtain the order parameter distribution $P(\Delta\phi)$ (Fig.~\ref{Fig4}c-bottom; red curve with open square symbol).
We note that bidispersity alone can drive MPS in the system, which is also evident from the sample snapshot shown in Fig.~\ref{Fig4}d for the case of $\epsilon_{HC}=0$ and $\Lambda=0$.  
This phase separation and the localization of the bigger beads (HC) at the surface of the cavity are driven by the depletion forces \cite{AsakuraJChemPhys1954, VrijPAC1976}.
Next, we focus on the effect of the enzymatic activity on this bidisperse setting keeping $\epsilon_{HC}=0$.
In the presence of enzymatic activity, we see MPS phenomenology with similar characteristic features observed for the MdDAM case, viz., asymmetric profile of $P(\Delta\phi)$ (Fig.~\ref{Fig3}c) and wall-like organization of EC regions (Fig.~\ref{Fig3}d) with the associated negative nematic order of the AA bonds (Fig.~\ref{Fig3}e). 
To compare the obtained MPS configurations for BdDAM with the corresponding MdDAM case, we calculate density-density cross-correlation between the two models (shown for A beads in Fig.~\ref{Fig4}f; also see Methods). 
The results imply a strong correlation between two model configurations under the action of enzymatic activity.

We also investigate the case in the presence of all\textemdash enzymatic activity, HC affinity, and bidispersity.
Even with the HC affinity, the tendencies similar to those mentioned right above are retained (Fig.~\ref{Fig4}c\textendash f; $\epsilon_{HC}=6$).
Therefore, we conclude that enzymatic activity affects the phase separation phenomenology in a similar way for both the monodisperse and the bidisperse setting.
However, we note also that there is still a visible difference in the MPS morphology due to bidispersity (Fig.~\ref{Fig4}d).


\section{Discussion} \label{discussion}
In summary, we have investigated the role of Topo-II in interphase chromatin organization using a random copolymer model with coarse-grained blocks representing euchromatic and heterochromatic regions, 
where Topo-II drives the system out of equilibrium. 
We noted that Topo-II has an intrinsic ability to microphase separate the chromatin.
To understand the underlying mechanism of this phase separation, we studied a simplified equilibrium polymer model as well as a simplified mean-field framework. 
These studies suggest that transient phantomness of a subsection of polymer due to Topo-II activity can drive this phase separation. However, in spite of being the essential mechanism for phase separation, it does not explain the characteristic wall-like organization of the EC regions that emerge due to precise mechanistic scheme of Topo-II activity.
Further, exploiting our polymer model, we show that bidispersity of chromatin due to different sizes of epigenetic marks affects its MPS morphology, however, the characteristic features of Topo-II induced MPS survives there.

Recently, chromatin organization has been investigated extensively in the context of phase separation, and several different mechanisms and models have been proposed. 
In ref.~\cite{SasaiBioRxiv2021}, repulsion-driven phase separation was proposed using polymer physics model simulations. 
It was assumed that the chromatin configuration in EC regions is looser and more flexible than in HC regions which allowed for the overlap between the monomers representing EC regions and led to the phase separation of the monomers representing EC and HC regions.
Our equilibrium self-avoiding-phantom-polymer model may be regarded as a limit of their repulsive-interaction-only polymer model.

In ref.~\cite{ShenoyBioRxiv2021}, the authors developed a field-based model simulating HC formation which incorporates the kinetics of methylation and acetylation 
in order to clarify the impact of the changes in histone methylation status on chromatin condensation.
They found that the methylation/acetylation reactions lead to interconversion of the EC and HC phases, and it provides more HC-EC interfaces. 
For the simulations performed in the present paper, we have used histone marks to designate EC and HC regions, and hence they are not changing in time. However, kinetics of histone mark alteration can induce phase separation. 
A possible future direction may be to integrate such reaction kinetics of histone modification into our polymer-based model and study how it can interplay with the enzymatic activity-induced phase separation in determining the HC-EC interface property. 

Our simulations as well as the aforementioned works focused on the EC- and HC-phase separation  in the nucleus. On the contrary, ref.~\cite{SafranELife2021} investigated the chromatin-aqueous phase separation. They found near-surface organization of the entire chromatin content. 
In our studies, we assumed high packing-fraction situation which does not allow for such chromatin-aqueous phase separation. 
However, we expect that our model will reproduce near-surface organization of both HC and EC if we extend it by considering a larger cavity, incorporating chromatin-lamin interactions, and tuning parameters like a polymer in bad solvent case (i.e., setting a strong inter-bead attraction). 

We saw that Topo-II induced microphase separation causes the wall-like appearance of EC domains. 
This is unlikely to be explained by another mechanism of phase separation proposed for chromatin organization in literature \cite{GautamBiophysJ2020, GautamNAR2014}, which relies on inhomogeneous effective-temperature.
Inhomogeneous temperature models are essentially out-of-equilibrium whereas, in the mechanism which we propose, the microphase separation itself happens even in equilibrium as suggested by the self-avoiding-phantom-polymer model study.
Therefore, our study highlights the importance of mechanistic models to understand the influence of out-of-equilibrium biophysical processes in chromatin organization. 

In conclusion, our study highlights the critical role of enzymatic activity in determining spatial features of eu- and heterochromatin architectures.
In general, there are a number of situations where heterochromatin architecture changes depending on the state or condition of a cell.
For example, aging correlates with the heterochromatin architecture \cite{LiEpigenetics2012, BohrExpMolMed2020}. 
Also, in aging, activity of various enzymes is known to  undergo  profound  changes  with  cell  state  modifications \cite{AdelmanBioscience1975, BeckerScience2021}.
A part of aging-associated alteration of heterochromatin architecture might be attributed to the variation of enzymatic activity.
Furthermore, alteration of heterochromatin architecture is observed for other cell state modifications like cell differentiation and under external forcing, although less is known about variation of enzymatic activity in those cases \cite{MeshorerDevCell2006,TalwarBiophysJ2013,DamodaranMBoC2018}.
Our finding suggests that further experiments focusing on the correlations between enzymatic activity and chromatin organization would provide hints to find out the mechanisms of such alteration of heterochromatin architecture and hence the cell state-specific genome regulation. 
 \\


\section{Methods} \label{methods}
\noindent{\bf Simulation setting for the monodisperse model.}
We simulate a $50.807$ Mbp long single chromatin packed within a spherical cavity of diameter $d_{ct}=1.4112    \:\mu\text{m}$. The size of the chromatin territory (i.e., the spherical cavity) was chosen to comply with a typical density of human diploid genome where $6.2$ Gbp DNA is packed within a nucleus of diameter $7$ $\mu$m. This single chromatin is mimicked by a bead-and-spring model comprising $N=20992$ equal-sized beads. Therefore, each bead represents $2.4203$ kbp chromatin. We assume nucleosomes as spheres of diameter $d_{nucleosome}=22$ nm (histone octamer core plus the linker DNA) containing $200$ bp of DNA. We further assume close compaction of nucleosomes within the beads A and B, such that $d_A=d_B= d_{nucleosome}\times(number \:of \:nucleosomes \:per \:bead)^{1/3}$. This determines the diameter of the beads as $d_A=d_B=50.5086$ nm for the monodisperse model.

In our active polymer model, we keep $\lambda_{an}=0.00167 \tau^{-1}$ and $\lambda_{an}=0.05\tau^{-1}$ fixed, $\tau$ being the unit time in our simulation, and the rate $\lambda_{ra}$ is treated as a simulation parameter. 
\\


\noindent {\bf Simulation setting for the bidisperse model.}
We set sizes of the beads, A and B, same as the mean sizes of the histone mark clusters, H3K9ac and H3K27me, respectively; therefore, $d_A=46 \:\text{nm}$ and $d_B=61 \:\text{nm}$ \cite{XuCellRep2018}.

To obtain a biologically relevant composition parameter (i.e., $\phi_A$ in our model), we extracted the radial distribution function (RDF, $g(r)$) data for the histone marks H3K9ac and H3K27me3 from ref.~\cite{XuCellRep2018}. Those RDF's were calculated by averaging over several two dimensional segments of the captured microscopy images. We calculated $m_{A} = \int_{segment} 2\pi r g_{A}(r) dr / \int_{0}^{d_{A}/2} 2\pi r g_{A}(r) dr$ (similarly $m_B$) and obtained 
the volume fraction of the A beads as $\phi_A=m_A V_A / (m_A V_A + m_B V_B) = 0.3544$.

We simulate the bidisperse model with the above mentioned bead sizes and volume fraction using the length of the polymer $N=17664$ that keeps the total DNA content (in bp) same as the monodisperse model. \\


\noindent{\bf Simulation units.}
We set $d_{ct} = 12 \:\ell$ that gives us the simulation unit (s.u.) of length as $\:\ell=117.6$ nm. Our model is simulated at a physiological temperature $T = 310$ K, and we consider energy unit as $e = 1 \;k_BT = 4.28$ pN$\cdot$nm. The frictional drag on monomers is approximated by Stokes' law, and the corresponding viscosity (of nucleoplasm) is assumed to be $1.5$ cP \cite{LiangJBiomedOpt2009}. Considering the nucleoplasmic viscosity as unity in simulations, we get the simulation time unit $\tau = 0.57$ ms. \\


\noindent{\bf Brownian dynamics.}
The position ${\bm x}_i$ of the $i^{th}$ bead is updated by integrating 
\begin{eqnarray}
\partial_t {\bm x}_i = - \frac{1}{6\pi\eta(d_i/2)} \partial_{\bm x}H_i + \sqrt{\frac{2k_BT}{6\pi\eta(d_i/2)}}{\bm \zeta},
\label{EqBrown}
\end{eqnarray}
where ${\bm \zeta}$ represents a univariate white Gaussian noise with zero mean, and $\eta$ represents the nucleoplasmic viscosity. $H$ comprises the following energetic costs\textemdash
\begin{itemize}
    \item Stretching energy between two beads, connected along the polymer and separated by $r$, is given by \\
    $-\frac{1}{2} k r_0^2 ln\left[1-\left(r/r_0\right)^2\right]$, where we choose $k=22 e$, $r_0=\sum_{i\in connected \: beads} d_i$.
    \item Confinement energy is approximated by the interaction potential between a wall and a star polymer with functionality $2$ (therefore, a linear polymer) \cite{JusufiJPCM2001, CamargoJCP2009}. For separation $r \leq \ell$ between the centre of the $i^{th}$ bead and the wall, the energy is given by $4\times2^{3/2}\left[ -ln(r/R_s)-0.0420(r^2/R_s^2-1)+0.8072 \right]$ and $4\times2^{3/2}\times 2.2550 \: erfc\left(2r/d_i\right)$, respectively, for the radius of corona $R_s=0.325 \:d_i \geq r$ and $R_s<r$.
    \item Volume exclusion potential between two beads separated by $r \leq \ell$ is given by $\epsilon_{vex}exp\left(-\alpha_{vex}r^2\right)$. The parameters are chosen such that the minimum of (volume exclusion + stretching) potentials occurs at $r_m=d_i$ for the monodisperse model and $r_m=\sum_{i\in interacting \: beads} d_i/2$ for the bidisperse model. We use $\epsilon_{vex}=8 \:e$ and $\alpha_{vex}=7.9585 \:\ell^{-2}$ for the monodisperse model. For the bidisperse model, we choose $\left( \epsilon_{vex}, \alpha_{vex}\right)$ as  $\left(6.6539, 9.5686\right)$, $\left(11.507, 5.5330\right)$, $\left(8.9153, 7.1414\right)$ in s.u. for the pairs AA, BB, and AB, respectively. For all these choices, we follow $\alpha_{vex} r_m^2 exp\left(-\alpha_{vex} r_m^2\right) = 2kr_m^2/3\epsilon_{vex}=constant$.
    \item HC affinity potential is taken to be $\epsilon_{HC}r^2exp\left[     -\alpha_{HC} \left\{d_B-1/(\alpha_{HC}d_B)-r\right\}^2\right]$ for a separation $r\leq\ell$, where we choose $\alpha_{HC}=100 \:\ell^{-2}$.
    \item In the transient attraction state of enzymatic activity, Topo-II pulls two A beads close to each other with a potential $-8exp\left(-7.9585 r^2\right) e$ for a separation $r\leq\ell$.
\end{itemize}

Equation~\ref{EqBrown} is integrated over time in Euler method, where we use a disctretized time step $10^{-4}\tau$. The simulation are done using lab-developed codes where we use CUDA to exploit GPU acceleration and OpenMP for CPU parallelization. We start the simulations from  equilibrated ideal chain configurations confined within the spherical cavity. The system is annealed for a time span (typically, $1.15$ s) by which the mean square displacement of a bead is more than the radius of the cavity \cite{MirnyPNAS2018}. Next, simulations are executed for the same duration as the annealing time, and numerous snapshots are stored. The results presented in the paper are obtained by analyzing such snapshots from at least $4$ different realizations for each set of parameters. \\


\noindent{\bf Quantification of phase separation.}
The whole cavity is divided into cubic grids of linear size $\ell$, and a parameter $v_i=(volume \; of \; i^{th} \; grid \; accessible \; to \; the \; beads)/\ell^3$ is calculated for each grid. The distribution of the order parameter $\Delta\phi$ is defined as 
\begin{eqnarray}
P\left(\Delta\phi\right) = \Big\langle \left[\sum_{i\in grids} v_i \delta\left(\Delta\phi-\Delta\phi_i\right)\right]/\left[\sum_{i\in grids}v_i\right] \Big\rangle_{snapshots},
\end{eqnarray}
where $\delta$ indicates Dirac delta function.
The Binder cumulant and the moment coefficient of skewness of $\Delta\phi$ are defined as $1-\langle(\Delta\phi)^4\rangle_P/3\langle(\Delta\phi)^2\rangle_P^2$ and $E\left[\left(\Delta\phi-\langle\Delta\phi\rangle_P\right)^3\right] / \left\{E\left[\left(\Delta\phi-\langle\Delta\phi\rangle_P\right)^2\right]\right\}^{3/2}$, respectively. Here, $E[\cdot]$ signifies the expectation operator. \\


\noindent{\bf Mean-field framework.}  
We consider a lattice space of size $M$ containing $M_{A^\prime}$ number of A$^\prime$ and $M_B$ number of B beads, such that $M=M_{A^\prime}+M_B$. Two A$^\prime$ beads can overlap to form a doublet (D) leaving an empty (E) site. We define the doublet fraction $\alpha = (number \: of \: D's)/M_{A^\prime}$, and therefore $0 \leq \alpha \leq 1/2$. Defining the volume fraction of B in this lattice model as $\Phi_B = M_B/M$, we can write the volume fractions of A$^\prime$, D, and E as $(1-\Phi_B)(1-2\alpha)$, $(1-\Phi_B)\alpha$, and $(1-\Phi_B)\alpha$, respectively. Given this setting, we can write the free energy density of the system as 
\begin{eqnarray}
f\left(\Phi_B,\alpha,\{\epsilon\}\right) = && \Phi_B \ln\Phi_B + \left(1-\Phi_B\right) \ln\left(1-\Phi_B\right) \nonumber \\
&&+ \left(1-\Phi_B\right)\left[ 2\alpha\ln\alpha + \left(1-2\alpha\right)\ln\left(1-2\alpha\right) \right] + f_{int}\left(\Phi_B,\alpha,\{\epsilon\}\right),
\label{free_ene_full}
\end{eqnarray}
where $\{\epsilon\}$ represents the pair interaction strengths among A$^\prime$, D, E, and B, and $f_{int}$ stands for the total interaction energy. We set $\epsilon_{BB}=\epsilon$, $\epsilon_{A^\prime B}=\epsilon$, $\epsilon_{DB}=2\epsilon$, $\epsilon_{A^\prime D}=\epsilon$, $\epsilon_{DD}=2\epsilon$, and rest of the pair interactions are set to zero. This choice of the interaction parameters gives us
\begin{eqnarray}
f_{int}\left(\Phi_B,\alpha,\{\epsilon\}\right) = \epsilon \left[ c(\alpha) \Phi_B^2 - 2c(\alpha) \Phi_B + \left( c(\alpha) + \frac{1}{2} \right) \right],
\label{free_ene_int}
\end{eqnarray}
where $c = -\alpha^2 + \alpha - 1/2$. Minimizing $f$ with respect to $\alpha$, we obtain the critical doublet fraction $\alpha^*\left(\Phi_B,\epsilon\right)$, and thereby we obtain the reduced free energy density $f^*\left(\Phi_B,\alpha^*,\epsilon\right)$. 
Note that we have considered only repulsive interactions among the lattice-pairs, and therefore, we call $\epsilon$ a pair repulsion parameter. \\


\noindent{\bf Nematic order parameter.}
The whole spherical cavity is gridded into cubic localities with lateral dimension $\ell$ ($=d_{ct}/12$ and $>2d_{A,B}$). We define a specific type of bond (i.e., AA or BB) as $\bm{u}=\bm{x}_{i+1}-\bm{x}_i$, $i \in [1, N]$. Given multiple ($>4000$) snapshots at equally separated time points for a realization, we count the total number of specific type of bonds $q_j$ in the locality $j$. We construct the local nematic tensor $\bm{Q}_j= \left( 3 \bm{u}_j \otimes \bm{u}_j - \bm{I} \right) / 2$ and diagonalize it. The largest eigen value of the $\bm{Q}_j$ is defined as the local nematic order parameter $S_j$. Grid wise local nematic order parameters are shown in Supplementary Fig.~3 for two sample cases. Note that the confinement induces a local nematic order of bonds near the surface, but we are interested to see order emerging due to enzymatic activity. So, we calculate the mean local nematic order parameter of AA bonds as $S = \sum_{j\not\in {surface}} q_j S_j / \sum_{j\not\in {surface}} q_j$. Here, we consider the outer most spherical shell of width $\ell$ as the surface region of the cavity.
\\


\noindent{\bf Segmentation of heterochromatic foci and analysis.}
We load the coordinates of B's on OVITO \cite{Ovito}, an open visualization tool, and use its cluster analysis modifier. Two B's separated by less than or equal to the bead size, $d_B$, are considered to be the members of the same cluster. Any cluster comprising atleast $(2\times block \:size)=8$ B's are considered as heterochromatic focus, otherwise neglected as noise.  

To quantify the size and the shape of the segmented heterochromatic foci, we calculate gyration tensor of individual focus, defined as $G_{mn} = \frac{1}{n_B} \sum_{i=1}^{n_B} r_m^{(i)} r_n^{(i)}$, where $r_m^{(i)}$ is the $m$-th Cartesian coordinate of the member $i$ of the $n_B$ B's forming the focus in its centre-of-mass frame. Diagonalizing ${\bm G}$, we obtain three eigen values, $\lambda_m^2$, $m = x, \:y, \:z$, along three principal axes of the focus. The radius of gyration of the focus is given by $\sqrt{\sum_{m = x, y, z} \lambda_m^2}$, and the shape anisotropy is defined as $\kappa^2 = \frac{3}{2} \frac{\sum_{m = x, y, z} \lambda_m^4}{\left( \sum_{m = x, y, z} \lambda_m^2 \right)^2} - \frac{1}{2}$. $\kappa^2$ will be zero for a spherical focus and unity when all the member-B's will fall on a straight line. \\


\noindent{\bf Surface localization profile calculation.}
We consider the outer most spherical shell of width $\ell$ as surface region. 
We calculate the ratio of the number of B's to the total number of beads in that shell, and average over multiple snapshots and realizations to obtain $\phi_{B, surface}$. \\


\noindent{\bf Density-density cross correlation.}
To compare the configurations obtained for MdDAM and BdDAM, we grid the whole cavity into cubic localities of lateral size $\ell$, and calculate the local density of A's. For a given snapshot of a specific model, if $c_{j}$ be the density at locality $j$, and $\bar{c}$ is the mean density therein, then we calculate the cross correlation as $\langle \frac{\sum_j\left[ \left( c_{j, MdDAM} - \bar c_{MdDAM} \right) \left( c_{j, BdDAM} - \bar c_{BdDAM} \right) \right]} {\sqrt{ \sum_j \left( c_{j, BdDAM} - \bar c_{BdDAM} \right)^2 \sum_j \left( c_{j, BdDAM} - \bar c_{BdDAM} \right)^2}} \rangle_{pair \;of \;snapshots}$.  


\section*{Acknowledgments}
We thank Andrew Wong from Mechanobiology Institute (MBI) science communication core  for editing the manuscript and MBI computational core for supporting us about computer-related research activities. We also appreciate Yuting Lou for valuable discussions.
This research was supported by Seed fund of Mechanobiology Institute (to JP, TH) and Singapore Ministry of Education Tier 3 grant, MOET32020-0001 (to GVS, JP, TH) and JSPS KAKENHI No. JP18H05529 and JP21H05759 from MEXT, Japan (to TS).

\subsection*{Author contributions}
R.D., G.V.S., J.P. and T.H. conceptualized the work, 
R.D., T.S., J.P. and T.H. constructed theory,
R.D. performed all the numerical simulations and visualized the results,
J.P. and T.H. supervised, 
R.D. and T.H. wrote the original draft, and 
all authors reviewed.

\subsection*{Competing interests}
The authors declare no competing interests.

\subsection*{Data and availability}
All data are available in the main text or the supplementary materials.

\newpage

\newpage
\renewcommand{\figurename}{Supplementary Figure}
\setcounter{figure}{0}
\begin{figure}[t]
\includegraphics[width=0.9\linewidth]{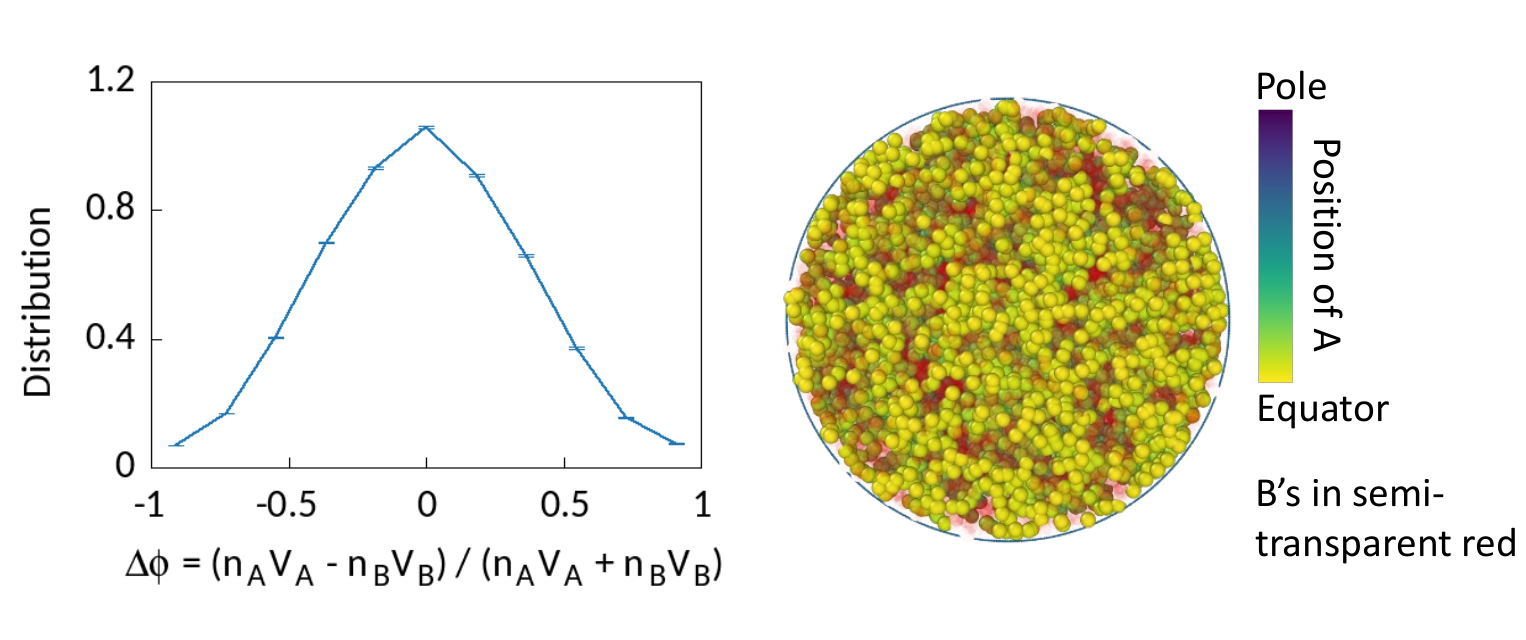}
\caption{ \footnotesize
{\bf Topo-II driven phase separation is not due to the transient attraction between A beads.}
Our monodisperse model shows strong phase separation for the heterochromatin affinity $\epsilon_{HC}=0$ and enzymatic activity $\Lambda=0.03094$. 
We mimic the transient attraction state of the enzymatic activity ($\Lambda=0.03094$) by an effective attraction strength among the A beads. Multiplying enzymatic activity (which is roughly equal to the ratio of the time spent in the attraction and the repulsion state) with the maximum attraction strength ($8$ s.u.) during the transient attraction, we get an effective attraction strength among the A beads as $0.25$ s.u.. Simulation of our monodisperse model for $\epsilon_{HC}=0$, $\Lambda=0$, but in the presence of an attraction among the A beads with strength $0.25$ s.u. (potential used is similar to the heterochromatin affinity case) does not show any phase separation, as understood from the snapshot (hemisphere cut) and the order parameter distribution. The error bars in the distribution plot indicate standard deviations over realizations.
}
\label{SupFig2}
\end{figure}

\begin{figure}[t]
\includegraphics[width=\linewidth]{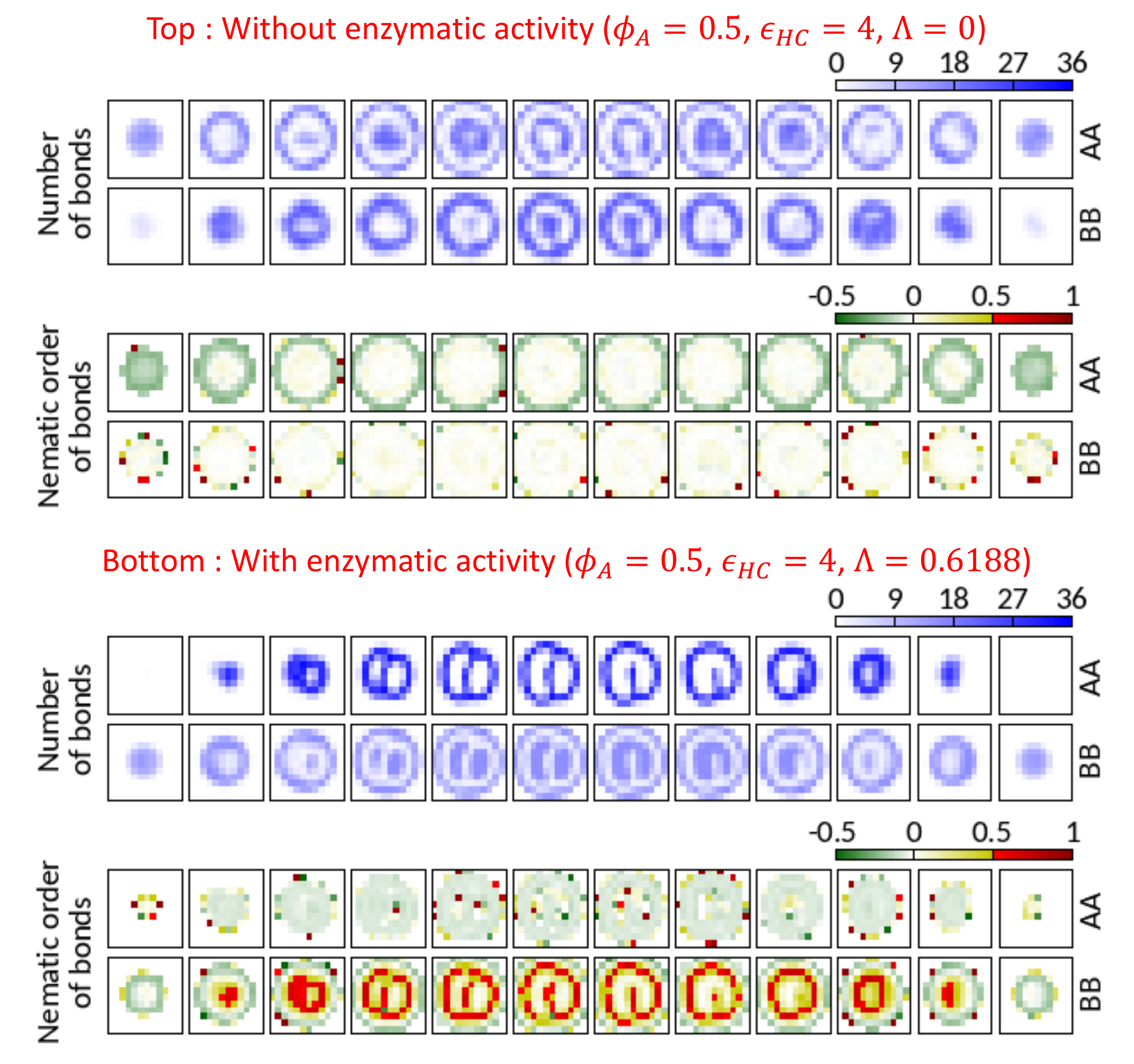}
\caption{ \footnotesize
{\bf AA bonds along the polymer are organized in planes of the walls formed by Topo-II.}
We count the average number of AA and BB bonds (springs) along the polymer in local grids ($\ell\times\ell\times\ell$ in size) and calculate the amplitude of the nematic order parameter of those bonds. Negative, zero, and positive order parameter values respectively imply planar ordering normal to the director, no ordering, and an ordering parallel to the director.  
Horizontally arranged twelve boxes represent twelve slices (thickness $\ell$) of the cavity. 
Top\textemdash In the absence of the enzymatic activity, no significant ordering is observed except (i) for AA bonds near the surface due to confinement constraint and (ii) in the grids where there are insufficient number of bonds to calculate order parameter.
Bottom\textemdash Due to the activity of Topo-II, A beads form wall (refer to the grids with high number of AA bonds) where the amplitude of the order parameter is negative. This suggests that the AA bonds along the polymer are organized in the plane of the wall with local directors normal to the wall (see the schematic in Fig.~3c). The strong positive amplitude for the BB bonds are due to sparse number of those bonds in the local grids.
}
\label{SupFig3}
\end{figure}
\begin{figure}[t]
\includegraphics[width=\linewidth]{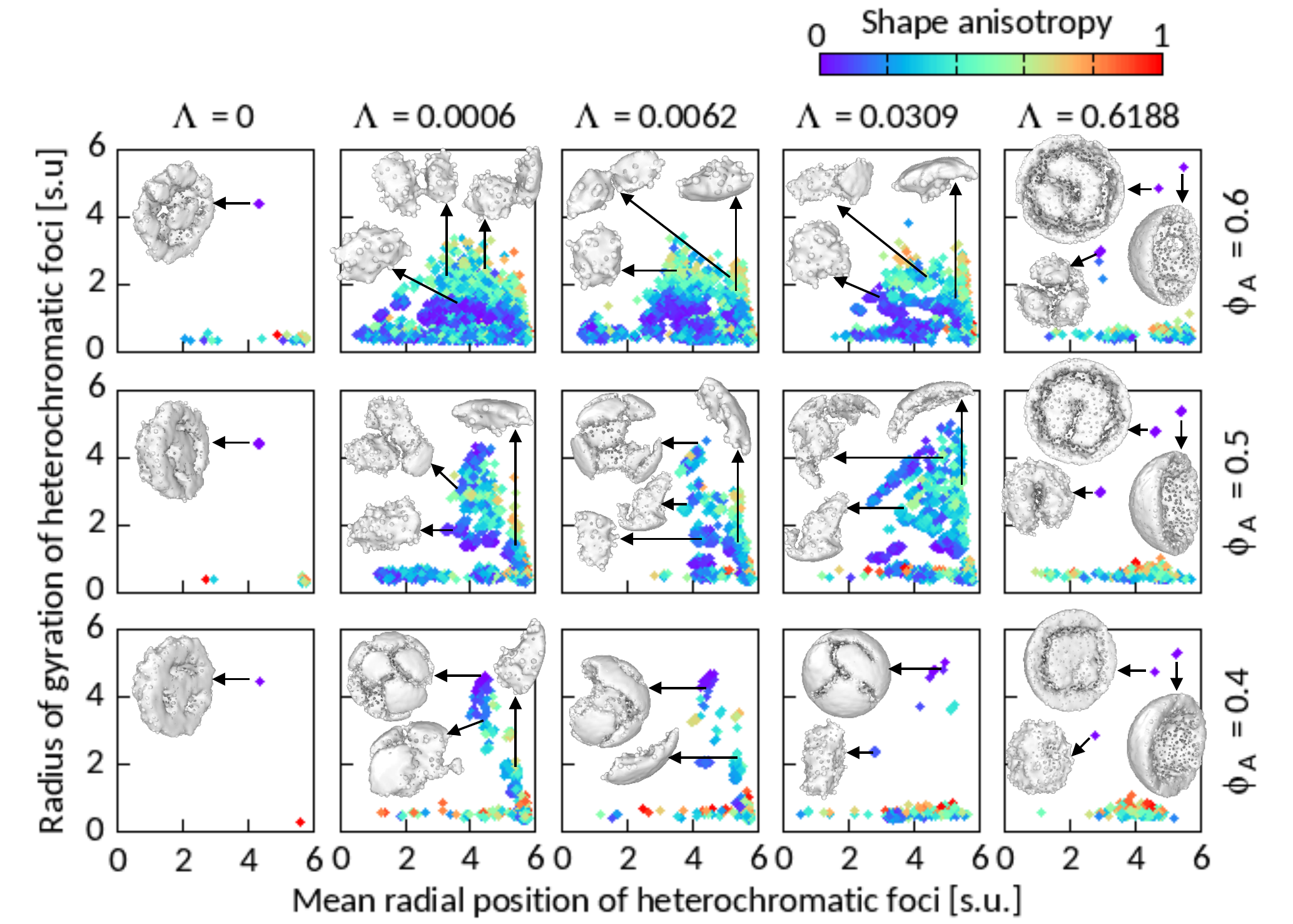}
\caption{ \footnotesize
{\bf Features of heterochromatic foci shown for $\epsilon_{HC}=4$.}
Radius of gyration of segmented foci are plotted against their mean radial position $r_{foci, i} = \left(\sum_{B\in i} r_{B}\right)/n_{B,i}$, $n_{B,i}\equiv$ number of B's forming foci $i$.
A few sample foci have been shown on the respective panels. The images of the foci are prepared using `ambient occlusion' and `construct surface mesh' modifiers of OVITO. Most of the foci for $\Lambda = 0$ and $0.6188$ are shown in hemisphere-cut view for better presentation. 
}
\label{SupFig4}
\end{figure}

\end{document}